# Diamond optomechanical crystals with embedded nitrogen-vacancy centers

**Jeffrey V Cady[1], Ohad Michel[1], Kenneth W Lee[1], Rishi N Patel[2], Christopher J Sarabalis[2], Amir H Safavi-Naeni[2] and Ania C Bleszynski Jayich[1]**

[1] Department of Physics, University of California Santa Barbara, Santa Barbara, CA 93106, USA
[2] Department of Applied Physics and Ginzton Laboratory, Stanford University, Stanford, CA 94305, USA

E-mail: ania@physics.ucsb.edu

**Abstract**

Hybrid quantum devices, in which disparate quantum elements are combined in order to achieve enhanced functionality, have received much attention in recent years due to their exciting potential to address key problems in quantum information processing, communication, and control. Specifically, significant progress has been made in the field of hybrid mechanical devices, in which a qubit is coupled to a mechanical oscillator. Strong coupling in such devices has been demonstrated with superconducting qubits, and coupling defect qubits to mechanical elements via crystal strain has enabled novel methods of qubit measurement and control. In this paper we demonstrate the fabrication of diamond optomechanical crystals with embedded nitrogen-vacancy (NV) centers, a preliminary step toward reaching the quantum regime with defect qubit hybrid mechanical devices. We measure optical and mechanical resonances of diamond optomechanical crystals as well as the spin coherence of single embedded NV centers. We find that the spin has long coherence times $T_2^* = 1.5$ μs and $T_2 = 72$ μs despite its proximity to nanofabricated surfaces. Finally, we discuss potential improvements of these devices and prospects for future experiments in the quantum regime.

## 1. Introduction

Hybrid mechanical systems, in which qubits are coupled to mechanical degrees of freedom, have emerged as a promising alternative and supplement to photonic architectures since phonons provide the advantages of reduced device size and crosstalk relative to photons of the same frequency while maintaining the ability to interact with a myriad of quantum systems. Recent experiments have demonstrated strong coupling between superconducting qubits and bulk acoustic resonators [1,2] and surface acoustic wave resonators [3]. This strong coupling enables the storage of quantum information in mechanical degrees of freedom [1] and preparation of phononic Fock states [2]. Another hybrid mechanical system that has shown promise utilizes strain coupling between diamond mechanical resonators and the negatively-charged nitrogen-vacancy (NV) center, which is an optically addressable defect in diamond that hosts a highly coherent spin [32]. Such NV-mechanical devices have enabled determination of the strain coupling strength to the NV center ground state spin [4,5] and orbital excited states [6] as well as demonstrations of mechanical methods of controlling the NV center spin [7,8,9] and orbital states [6,10]. Furthermore, mechanical coupling has been demonstrated using a variety of mechanical oscillators, including single crystal diamond cantilevers [4,5,6], bulk acoustic resonators [7,8], and surface acoustic waves [10]. However, in order to reach the strong coupling regime with NV centers, smaller mechanical resonators are required to achieve larger strain couplings, all while preserving the coherence of both the NV center and mechanical resonator. For the specific case of cantilevers and doubly-clamped beams these requirements prove challenging. Specifically, moving to smaller device sizes compromises the mechanical quality factor through increased clamping loss, and the strain profile of their flexural modes requires NV centers to be formed near the diamond surface in order to experience large strain, exposing them to surface-related decoherence mechanisms [24,25]. For many quantum applications, the basic requirement is realizing a cooperativity $C = \frac{g^2}{\Gamma \gamma} > 1$, where g is the strain coupling to the NV center ground state spin or orbital excited state due to zero-point motion of the mechanical resonator, $\Gamma$ is the decoherence rate of the NV center spin or orbital excited state, and $\gamma$ is the phonon decoherence rate.

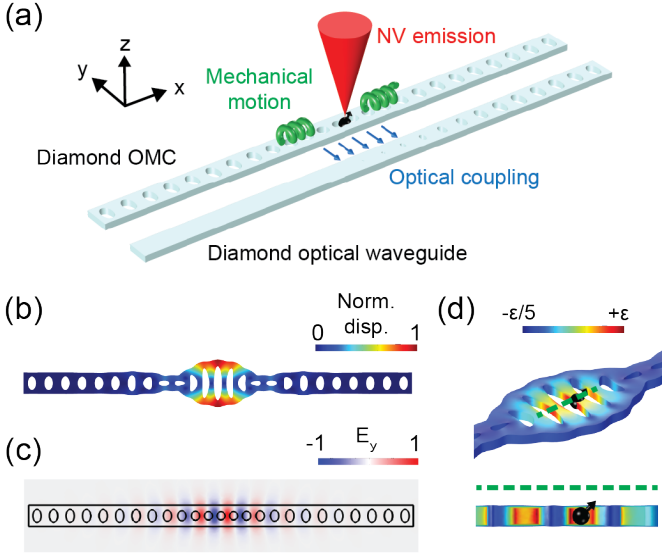

Figure 1: a) The system under study, in which an NV center is embedded near the center of a diamond OMC and interacts with the mechanical motion of the OMC via crystal strain. The NV is measured by collecting the light it emits with a confocal microscope, while the OMC is probed via coupling to an adjacent diamond optical waveguide. b) The normalized displacement of the fundamental "breathing" mechanical mode and c) the normalized electric field of the optical mode of the diamond OMC, simulated using the finite-element method (FEM). d) The $\varepsilon_{yy}$ component of the strain tensor in the device basis due to the displacement of the breathing mode, also simulated with FEM. The line cut shows the z-symmetric nature of the strain profile and the optimal placement of an NV in a large-strain region that is also isolated from the device surfaces.

In this work we demonstrate a step toward this requirement by designing and fabricating single-crystal diamond optomechanical crystals (OMCs) with stable embedded NV centers, shown by the diagram in figure 1a. We choose to work with OMCs to avoid the challenges associated with cantilevers and doubly-clamped beams, since OMCs host mechanical modes which are well-isolated from clamping points and which have strain profiles that are uniform throughout the thickness of the device, negating the requirement that the NV centers must be formed near the surface in order to experience high strain. Importantly, the NV centers in our devices are charge stable and exhibit long spin coherence times on the order of 100 μs, despite residing in thin, etched diamond structures.

We employ a one-dimensional OMC design that hosts colocalized and coupled optical and mechanical resonances. This architecture has experienced much success in silicon [11], silicon nitride [12], aluminum nitride [13] and, more recently, GaAs [14] and diamond [15]. The optomechanical coupling in these devices allows for laser cooling to their motional ground state [11] and remote entanglement of two OMCs [16], while their mechanical modes have been shown to be highly coherent, with mechanical quality factors exceeding $10^7$ [17]. OMCs are also capable of being coupled to propagating phonons in the form of surface acoustic waves [18] and phononic wire modes [19], indicating their potential versatility in future quantum information processing and communication applications.

## 2. Diamond optomechanical crystals

Diamond possesses several properties that make it a promising material for optomechanical crystals. Its large Young's modulus and low thermoelastic damping have been shown to enable high-quality factor mechanical resonators [20,21]. Furthermore, diamond possesses a large (5.5 eV) bandgap and high thermal conductivity, which should mitigate the effects of optical absorption that limit the effectiveness of laser cooling and driving of mechanical motion in silicon optomechanical crystals. Diamond also hosts a number of optically active defects, including the NV center and the negatively-charged Silicon-vacancy center, which are naturally interfaced with the mechanical motion of the OMC via crystal strain [22]. Previous demonstrations of diamond optomechanical systems can be found in [15,33].

Our OMC design consists of a single-crystal diamond nanobeam with a rectangular cross-section and a one-dimensional array of etched ellipses along its length and is fabricated using a diamond-on-insulator (DOI) technique. The rectangular cross-section is in contrast to the work in [15], which employs diamond OMCs with triangular cross-sections fabricated using an angled-etching technique. The rectangular cross-section enables important advantages, namely the z-symmetric strain profile of the fundamental "breathing" acoustic mode and the ability to fabricate two-dimensional phononic shields and couple to single-mode waveguide structures [19].

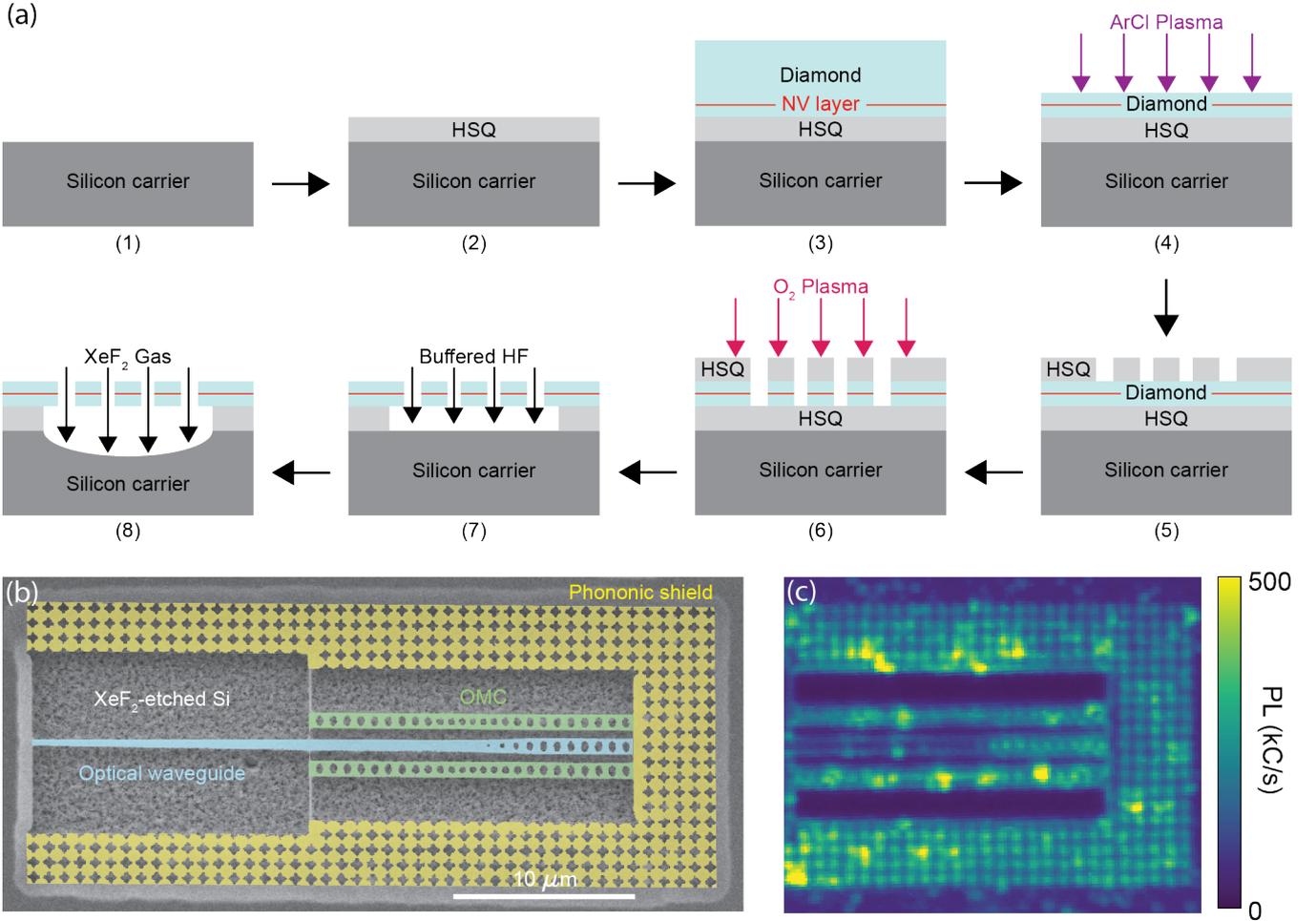

Figure 2: a) A simplified fabrication flow for diamond OMCs. 1) A silicon carrier chip is prepared and 2) coated with HSQ. 3) A ≈ 20 µm-thick diamond membrane is wafer-bonded to the HSQ above the curing temperature for HSQ. 4) The diamond is thinned to the device layer thickness of 200 nm using a series of ArCl inductively coupled plasma (ICP) etches. 5) HSQ is patterned on top of the diamond using electron beam lithography and then 6) used as an etch mask to define the devices in the diamond with an $O_2$ ICP etch. 7) The patterned devices are undercut with a buffered hydrofluoric acid dip, which selectively etches the HSQ bonding layer underneath the devices. 8) The devices are further undercut using $XeF_2$ gas, which isotropically etches the silicon carrier wafer. b) SEM micrograph of a pair of diamond OMCs on either side of a diamond optical waveguide. The devices are surrounded by a cross pattern phononic shield. c) A scanning confocal micrograph of a device similar to that shown in a). Single NVs can be discerned in the area surrounding the phononic shield but scattering from the etched holes and sidewalls makes them difficult to see inside the OMCs.

We engineer the OMC optical resonance to be ~200 THz to facilitate integration with mature 1550 nm telecommunication technologies. OMCs with this optical frequency also host mechanical resonances with frequencies of a few GHz. We choose an OMC thickness of 200 nm, which provides a balance between the competing values of optical quality factor, for which a thicker OMC would be optimal, and strain, which increases for thinner OMCs. We then follow the design principles outlined in [23] in which "mirror" unit cells at either end of the OMC adiabatically transition to a "defect" unit cell at the center through slight changes in the shape and spacing of the etched holes in the region between the defect and mirrors. The mirror cells host incomplete, symmetry-dependent acoustic and optical band gaps around the breathing acoustic mode (figure 1b) and fundamental transverse-electric (TE) optical mode (figure 1c) frequencies. Further acoustic isolation is provided by a phononic shield surrounding the OMCs, which hosts a full phononic bandgap around the breathing mode resonance frequency. To optimize our OMC design for optical quality factor and strain, we simulate the optical and mechanical resonances of the OMC with the finite element method (FEM) software COMSOL and use a genetic algorithm to search the design parameter space and maximize the product $Q_o\varepsilon_0$, where $Q_o$ is the optical quality factor for the fundamental TE optical mode and $\varepsilon_0$ is the strain due to zero-point motion of the mechanical resonance.

The strain profile of the breathing acoustic mode for our rectangular cross-section devices is constant throughout the thickness of the device, allowing an embedded NV center to experience high strain even if it is equidistant from the top and bottom surfaces of the OMC, isolating the NV center from surface-induced noise [24,25]. Thus, as shown in figure 1d, the ideal placement of an NV center in one of our diamond OMCs based on the FEM simulated strain profile for the breathing mode would be approximately halfway between the defect hole and the adjacent hole and equidistant from the top and bottom surfaces of the diamond. This would place the NV in a high-strain region that is approximately 100 nm from any surface, providing isolation from surface-related decoherence mechanisms. At this location, we simulate the strain due to zero-point motion to be ~$10^{-8}$. Using the formalisms and measured stress coupling constants presented in [34,35] for the spin interaction and [36] for the orbital interaction, we estimate zero-point coupling strengths of ~200 Hz and ~10 MHz, respectively.

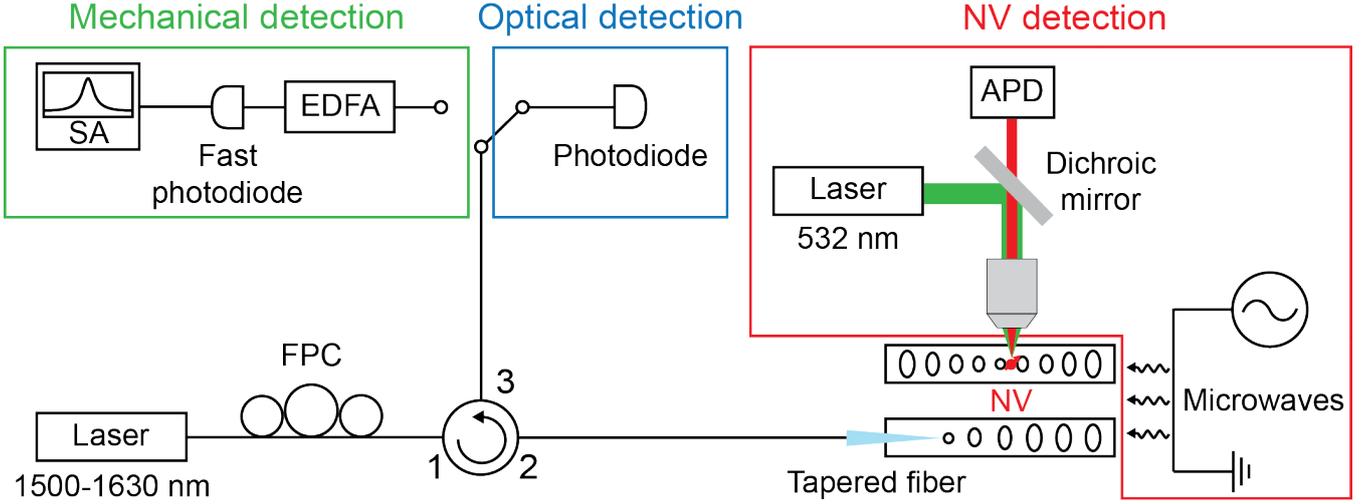

Figure 3: A combined NV and OMC characterization setup. To measure the optical response of the OMC, light from a tunable laser is sent through a fiber polarization controller (FPC) and a fiber circulator to a tapered optical fiber, which couples light into an on-chip diamond optical waveguide. The light reflects from the OMC and a patterned mirror at the end of the waveguide and re-enters the tapered fiber. The light goes once more through the fiber circulator before being measured with a DC photodiode. To measure the thermal mechanical motion of an OMC, the laser wavelength is set within an optical resonance but instead of being measured with a DC photodiode, is amplified with an erbium-doped fiber amplifier (EDFA) and sent to a fast photodiode which converts the intensity-modulated light into an electrical signal which can be measured with a spectrum analyzer (SA). NV measurements are carried out with a home-built scanning confocal microscope in which 532 nm light is used to excite the NVs and fluoresced red photons are measured using an avalanche photodiode (APD). Microwaves for NV spin manipulation are delivered via a wire bonded over the diamond.

### 3. NV center formation and device fabrication

Fabrication of our diamond OMCs with embedded NV centers begins with formation of NV centers in a 20 μm-thick single crystal diamond plate. $N^{14}$ is implanted with a dose of $3 \times 10^9$ cm$^{-2}$ at an energy of 19 keV and tilt of 7 degrees, which ultimately yields a ~100 nm deep layer of NV centers with a density of ~1 um$^{-2}$. Subsequently, the sample is annealed at 450 C in an $O_2$ atmosphere and next at 850 C in a $H_2$/Ar atmosphere. In the final step, the sample is cleaned in a boiling perchloric, nitric, and sulfuric acid mixture to remove any graphitization incurred during the annealing as well as to oxygen terminate the surface.

Next, as shown in figure 2a, the diamond plate is mounted NV side down to a silicon carrier chip with Hydrogen silsesquioxane (HSQ) and cured in a wafer bonder at 420 C and with 80 mBar of bonding pressure. The diamond is then thinned to the device thickness of 200 nm using an ArCl inductively coupled plasma (ICP) etch. A grid of OMCs with in-plane dimensions scaled in the range ±7.5% from the original design is patterned in another layer of HSQ on top of the diamond using electron beam lithography. This pattern is then transferred into the diamond with an $O_2$ ICP etch. A buffered hydrofluoric acid dip then undercuts the OMCs by etching the underlying HSQ. Finally, the OMCs are further undercut by exposure to $XeF_2$ gas, which etches the silicon underneath the OMCs in order to minimize leakage of the OMC optical field into the silicon substrate. A completed device is shown in figure 2b.

### 4. Optomechanical characterization

We measure the optical and mechanical resonances of our OMCs using a room-temperature fiber setup, as shown in figure 3. For measurement of optical resonances, light is sent from a wavelength-tunable laser (1500-1630 nm) through a fiber polarization controller (FPC) to match the polarization of the light to that of the OMC TE optical mode, then through a fiber circulator. To couple the light onto the diamond chip, we employ a tapered optical fiber setup such as in [26], in which light is transferred from a tapered optical fiber into an on-chip diamond optical waveguide (which is centered between two nominally identical OMCs, as shown in figure 2a). The incident light can then evanescently couple into the OMCs when on resonance with the optical mode hosted by the OMC or continue down the waveguide and reflect off mirror cells which are designed to be identical to those in the OMCs. The reflected light then couples back into the tapered fiber and gets circulated to a photodiode, where the DC power of the reflected light is measured. Light that interacts with the OMC before leaking back into the waveguide gains a phase relative to the light that reflects off the waveguide mirror; an optical resonance is thus identified by a dip in the reflected power spectrum at the OMC optical resonance wavelength due to interference between these two optical paths. Figure 4a shows such a spectrum with an optical resonance at 1564.2 nm and a quality factor of 42,000. Since our optical resonances are not in the sideband-resolved regime for the 6 GHz OMC acoustic mode, to detect the mechanical mode, we set our laser to a wavelength within the bandwidth of the OMC's optical resonance. The mechanical motion of the OMC imprints phase fluctuations onto the light that interacts with the cavity, which then interferes with light that reflected off the waveguide mirror. The resulting intensity-modulated reflected optical signal is amplified with an erbium-doped fiber amplifier before being measured with a fast (12 GHz) photodetector and sent to a signal analyzer. Figure 4b shows the detected mechanical spectrum, which we associate with the breathing acoustic mode. A fit to this signal gives a mechanical quality factor of 118. This low mechanical quality factor, when compared to other devices in ambient conditions [15], could be due to leftover HSQ on the backside of the devices that was not completely removed during the undercut step (step 7 in figure 2a) or deviations in the fabricated device and phononic shield dimensions from the design dimensions.

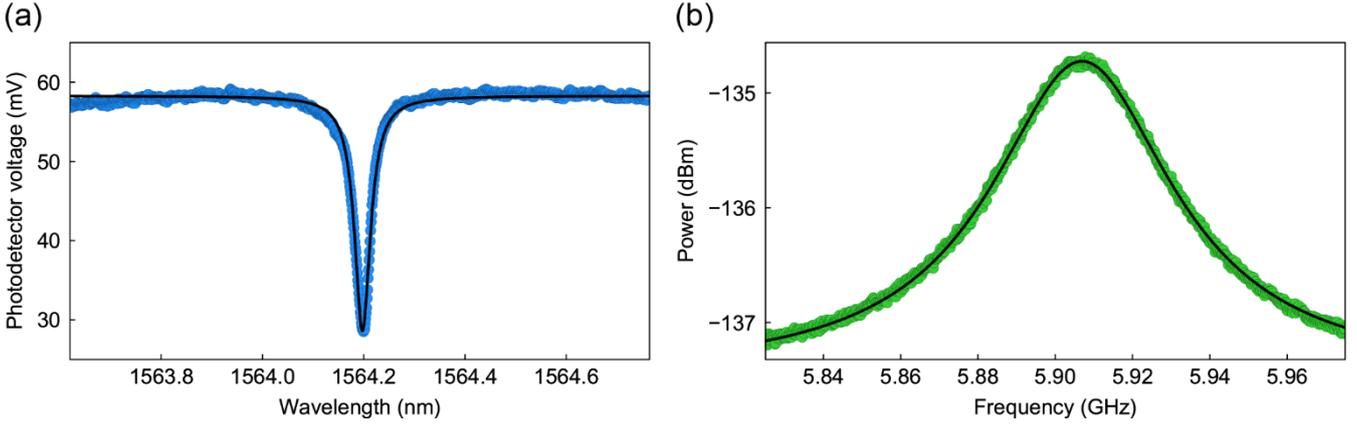

Figure 4: a) A typical optical resonance for a diamond OMC. A Lorentzian fit gives a quality factor of 42,000. b) The measured mechanical spectrum, which we associate with the OMC breathing mode. A fit to the signal gives a quality factor of 118.

## 5. NV center spin characterization

Preserving the spin properties of NV centers near nanofabricated surfaces is a perennial challenge. To characterize the ground state spin properties of the NV centers in our OMCs, we use a home-built confocal microscope, a simplified version of which is shown in figure 3. A fluorescence image of representative OMC devices is shown in figure 2c. We first detect the presence of NV centers near the center of our OMC using optically detected magnetic resonance (ODMR) of the NV center spin, in which the NV center is continuously excited with 532 nm light while being irradiated with microwaves. A dip in the photoluminescence of the NV center near 2.68 GHz (figure 5a) corresponds to a transition from the $m_s = 0$ spin state to the $m_s = -1$ spin state, with a small external magnetic field shifting the transition from the 2.87 GHz zero-field splitting. Three dips are present due to hyperfine coupling to the $N^{14}$ nuclear spin of the NV center. The ability to resolve hyperfine splitting is already indicative of well-preserved spin coherence.

To quantitatively characterize the spin coherence of this NV center, we perform microwave-driven Rabi, Ramsey, and Hahn echo measurements in which the NV center is prepared in the $m_s = 0$ ground state with a 532 nm laser pulse before undergoing the microwave pulse sequences called for by these measurements. For the measurement of Rabi oscillations, the NV center spin evolves under continuous exposure to microwaves resonant with the transition from the $m_s = 0$ state to one of the hyperfine sublevels of the $m_s = -1$ state for some amount of time before the spin-state is read out via spin-dependent fluorescence. Figure 5b shows persistent Rabi oscillations between these two states. Deviation of the signal from a cosine function is due to beating with the other hyperfine sublevels. For the Ramsey measurement shown in figure 5c, the spin is pulsed into a superposition of the $m_s = 0$ and $m_s = -1$ states with microwaves slightly detuned from the transition. The spin then freely evolves for $T = \tau$ before being projected back onto the $m_s = 0$ state and measured. The decay of the resulting oscillations, which are due to interference between the different hyperfine sublevels, is fit to $\exp(-(\tau/T_2^*)^2)$, where $T_2^*$ is the bare coherence time of the NV center spin. From the fit we extract $T_2^* = 1.5$ μs. In order to measure the NV center spin coherence when decoupled from slowly evolving noise sources, we perform a Hahn echo measurement. Microwaves resonant with the $m_s = 0$ to $m_s = -1$ transition place the spin in a superposition of the two states, after which the spin is allowed to evolve for $T = \tau$ before a refocusing pulse is applied and the spin is again allowed to evolve for $T = \tau$, for a total free evolution time of $2\tau$. The result of this measurement is shown in figure 5d. We associate the oscillations in the signal with hyperfine coupling to a proximal

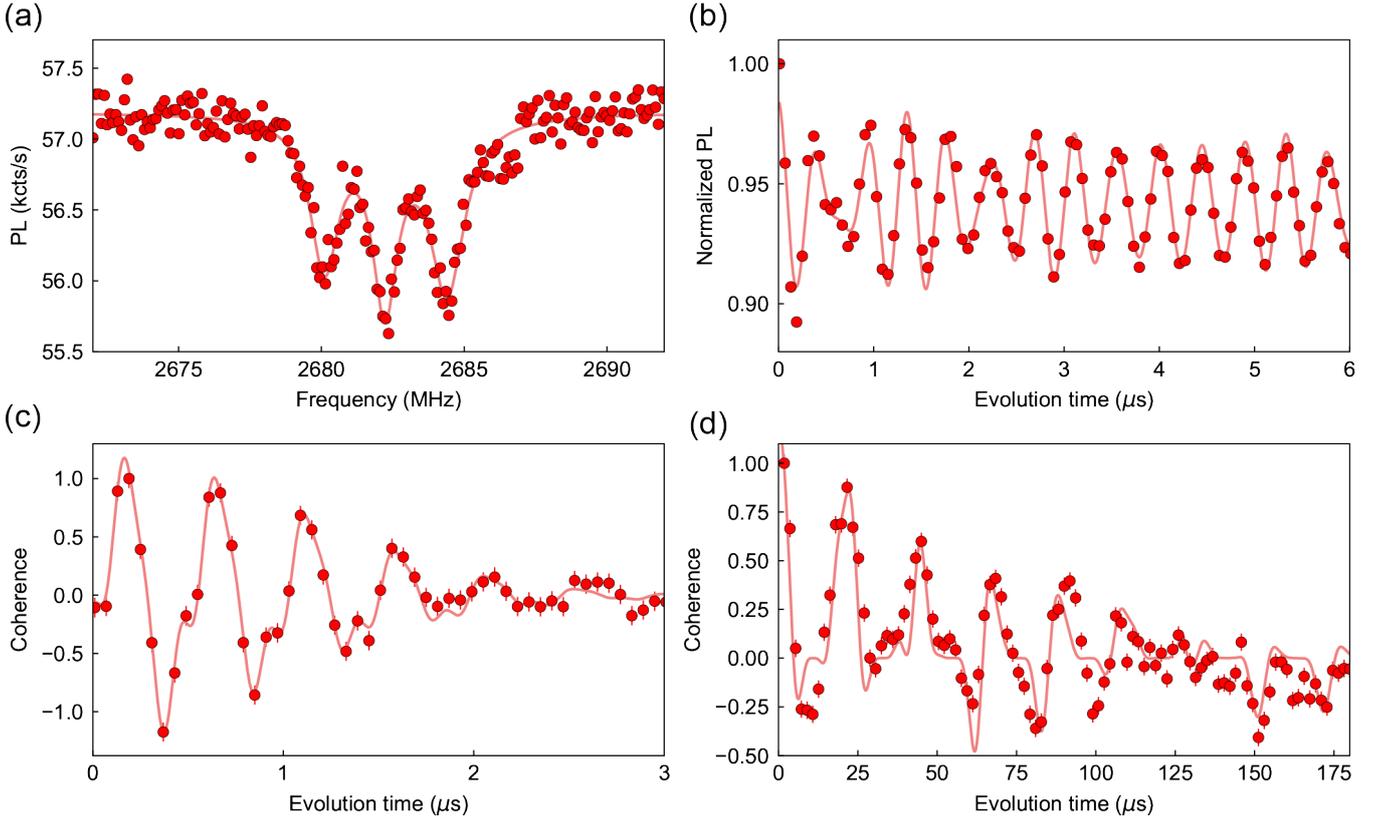

Figure 5: NV characterization measurements for an NV located near the center of an OMC. a) Microwave-driven optically-detected magnetic resonance measurement of the $m_s = -1$ spin state of the NV, in which the hyperfine sublevels due to the adjacent $N^{14}$ nucleus are resolved. b) Persistent, microwave-driven Rabi flopping between the $m_s = -1$ and $m_s = 0$ spin states of the NV. c) A Ramsey measurement of the qubit formed by the $m_s = 0$ and $m_s = -1$ spin states. A fit to the decay gives $T_2^* = 1.5$ μs. d) A Hahn echo measurement on the same states as in c). A fit to the signal that includes both the effects of a $C^{13}$ spin bath and a proximal $C^{13}$ nucleus gives $T_2 = 72$ μs. In c) and d) error bars indicate one standard deviation.

$^{13}$C nuclear spin as well as the Larmor precession of the $^{13}$C nuclear spin bath present in the diamond. The decay of this signal is fit to $\exp(-(2\tau/T_2)^2)$, where $T_2$ is the Hahn echo coherence time of the NV center spin. From the fit we find $T_2 = 72 \pm 6$ μs.

## 6. Discussion and outlook

In summary, we have demonstrated the fabrication of rectangular cross-section diamond optomechanical crystals with embedded NV center spins. Rectangular cross-section OMCs are particularly promising for hybrid NV-mechanical systems due to the z-symmetric strain profile of the fundamental breathing acoustic mode and the ability to fabricate two-dimensional phononic shields, which should allow for high mechanical quality factors. Our diamond OMCs host optical modes with quality factors of up to 42,000 and an acoustic breathing mode near 6 GHz with quality factors of ~100 in ambient conditions. Importantly, our fabrication process preserves the long spin coherence time of NV centers, which we demonstrate by measuring a coherence time of $T_2 = 72$ μs.

To reach the high cooperativity regime for coupling between an NV center spin and the mechanical motion of a diamond OMC, improvements to the mechanical quality factor and NV center spin coherence are necessary, in addition to larger engineered strain. For an OMC with a mechanical quality factor of $10^6$ in its ground state of motion, a zero-point coupling strength of ~1 kHz, and NV center spin $T_2$ of 10 ms [27], C > 1 becomes possible. To this end, future iterations of these devices will implement modifications to the design and fabrication procedure to ensure full removal of the HSQ from the backside of the OMCs as well as accuracy of the fabricated device dimensions in an attempt to increase the mechanical quality factor. Furthermore, by using diamond that has been grown with isotopically pure $^{12}$C and using a delta-doped nitrogen layer and electron irradiation to form NV centers [28], we hope to improve the NV center spin properties in our OMCs. If the high cooperativity regime is reached, experiments such as quantum state transfer between 1550 nm light and an NV center spin and phonon-mediated spin-spin interactions become realizable.

While the NV center spin properties have been preserved in our OMCs, future experiments at low temperature will be required to determine the properties of the NV center orbital excited state. For a zero-point coupling strength of 10 MHz, a mechanical quality factor of $10^6$, and an excited state linewidth of 100 MHz [29], C ≈ 10 for the orbital-phonon interaction. This high cooperativity would enable, for example, cooling of the mechanical resonator using phonon-assisted transitions to the NV center orbital excited state. A final phonon occupation $n_f \sim 1$ is possible using either a resonant or off-resonant cooling scheme, assuming optical Rabi frequencies of ~1 MHz and ~1 GHz, respectively [30]. Furthermore, the larger strain interaction strength in the excited state can be used to provide an enhanced spin-phonon interaction using phonon-assisted Raman transitions [10,31], indicating a potential alternative path to high cooperativity for NV spin-mechanical devices.

## Acknowledgements

This work is supported by an NSF-CAREER award (DMR-1352660) at UCSB, and joint work between Stanford and UCSB acknowledges support of NSF award QIS-1820938.